\pdfoutput=1
\documentclass[11pt]{article}

\usepackage[margin=1.1in]{geometry}
\usepackage{amsmath}
\usepackage{amssymb}
\usepackage{booktabs}
\usepackage{microtype}
\usepackage{lmodern}
\usepackage{graphicx}
\graphicspath{{../reports/figures/}{./}}
\usepackage[round]{natbib}
\usepackage[hidelinks]{hyperref}

\newcommand{\RPS}{\mathrm{RPS}}

\newcommand{\Prob}{\mathbb{P}}

\title{\bfseries Does a Structural Model Add Anything to the Closing Price?\\[4pt]
\large Calibrated forecasting, incremental information, and match leverage\\
in the Italian Serie A}

\author{Yannik Pitcan\\
\small \texttt{pitcany@gmail.com}}

\date{August 2026}

\begin{document}
\maketitle

\begin{abstract}
\noindent
Studies of association-football forecasting routinely report three-way accuracy
in the low fifties and present it as competitive with the betting market. We
argue that accuracy against a uniform benchmark answers the wrong question, and
that the question worth asking is whether a model carries information a
margin-free closing price has not already absorbed. We formalise that test as
the fitted weight in a logarithmic opinion pool and apply it to nineteen
complete Serie A seasons (7{,}220 matches, 2007--08 to 2025--26).

The answer is negative and stable. A Dixon--Coles model with tuned exponential
decay attains 53.4\% accuracy and RPS 0.1972 against the market's 0.1905; the
paired difference is $+0.0067$ ($95\%$ CI $[0.0046, 0.0088]$) and the market
wins in all seven test seasons. The fitted pooling weight on the structural
model is $0.000$, and the log-loss profile is monotone increasing in that
weight on the validation period and on the test period alike, so the result is
a genuine boundary solution rather than an optimisation artefact. To test
whether this reflects the noisiness of goals rather than the method, we refit
the identical machinery to shots on target, a pre-expected-goals proxy for
chance creation. That variant earns weight $0.35$ against the goals model---it
demonstrably carries information the goals model lacks---and $0.000$ against
the market. Two structural signals, each informative relative to the other,
both already priced.

Separately, we find that the structural model is \emph{better calibrated} than
the market on the home-win margin (calibration slope $0.995$ versus $1.103$)
while being clearly less sharp, so the market's advantage is discrimination
rather than honesty; studies reporting only accuracy cannot distinguish these.

The constructive consequence is that value lies not in a better outcome
forecast but in what is built on a calibrated one. We define \emph{match
leverage}, the change in a club's probability of achieving a season objective
between winning and losing a given fixture, give conditions under which it can
be estimated by post-hoc conditioning on a single simulation, and compute it
for ACF Fiorentina. The ordering inverts intuition: an away fixture against a
relegation rival carried $2.25\times$ the leverage of hosting the eventual
champions.

The paper also documents and corrects specific errors in an earlier study of
our own on the same problem.

\end{abstract}

\section{Introduction}
\label{sec:intro}

Association-football forecasting occupies an unusual position among applied
prediction problems: a strong, continuously updated, publicly observable
benchmark exists in the form of bookmaker prices. This has two consequences
that the applied literature has absorbed unevenly. First, any accuracy figure
must be read against both a lower bound (the outcome base rate, which home
advantage alone pushes well above uniform guessing) and an upper reference (the
margin-free market price). Second, and more importantly, a model's usefulness
is not established by scoring well; it is established by carrying information
the market has not already priced.

The second point is easy to state and easy to violate. A common design supplies
bookmaker odds to a classifier as input features and then compares the
resulting accuracy with the bookmaker's own. Under that design the model
consumes the market's forecast and is then credited for approaching it. The
appropriate reading of such a result is that the pipeline is discarding
information, not adding it.

\subsection{Contributions}

\begin{enumerate}
\item We formalise the incremental-information question as the fitted weight in
a logarithmic opinion pool with the de-vigged market price, and argue it should
be the primary reported quantity in this literature
(Section~\ref{sec:pooling}).

\item We report that weight to be $0.000$ for a well-specified Dixon--Coles
model over nineteen Serie A seasons, and establish that it is a boundary
solution rather than an optimisation artefact by tracing the full log-loss
profile on both validation and test data (Section~\ref{sec:pool-result}).

\item We rule out target noise as the explanation by refitting the identical
machinery to shots on target. That variant earns weight $0.35$ against the
goals model but $0.000$ against the market, showing that two structurally
different signals are both already priced (Section~\ref{sec:shots}).

\item We separate calibration from discrimination empirically, finding the
structural model better calibrated but less sharp than the market
(Section~\ref{sec:calibration}).

\item We define \emph{match leverage} and give the independence condition under
which post-hoc conditioning on a single season simulation is equivalent to
intervening on a fixture and resampling (Section~\ref{sec:leverage}).

\item We document and correct specific errors in an earlier study of our own on
this problem (Section~\ref{sec:errata}).
\end{enumerate}

\section{Related work}
\label{sec:related}

\paragraph{Goal-based structural models.} The modern lineage begins with
\citet{maher1982}, who modelled home and away goals as independent Poisson
variables with club-specific attack and defence parameters.
\citet{dixoncoles1997} identified the systematic misfit of the independence
assumption on low scores, introduced a dependence correction on the four cells
$(0,0)$, $(0,1)$, $(1,0)$, $(1,1)$, and added exponential down-weighting of
older matches. \citet{karlis2003} developed the genuine bivariate Poisson
alternative. Dynamic and Bayesian treatments followed:
\citet{rue2000} modelled time-varying strengths through a Bayesian dynamic
linear model, \citet{baio2010} introduced hierarchical shrinkage of club
parameters, and \citet{koopman2015} placed the bivariate Poisson in a
state-space framework. \citet{goddard2005} compared goal-based and
result-based regression formulations directly. \citet{hvattum2010} evaluated
Elo ratings as covariates, and---relevant here---benchmarked them explicitly
against market odds.

\paragraph{Market prices as forecasts.} \citet{forrest2005} compared
odds-setters with statistical forecasting models over several seasons and found
the former difficult to beat, with the gap narrowing over time.
\citet{shin1992,shin1993} derived the insider-trading model that underlies the
margin-removal method used here; \citet{strumbelj2014} evaluated competing
margin-removal schemes and found Shin's the best-performing practical choice.
This body of work establishes the market as a benchmark, but comparisons are
typically made on accuracy or on returns rather than by testing directly for
incremental information.

\paragraph{Scoring rules and calibration.} \citet{epstein1969} introduced the
Ranked Probability Score for ordered categorical forecasts;
\citet{gneiting2007} give the general theory of proper scoring rules.
\citet{constantinou2012} argued specifically that accuracy and other common
measures are inadequate for football forecasts because the outcome is ordered,
and established the RPS as the appropriate default. Calibration as a property
distinct from discrimination is treated by \citet{niculescu2005}; post-hoc
correction methods include Platt scaling \citep{platt1999}, isotonic regression
\citep{zadrozny2002} and temperature scaling \citep{guo2017}.

\paragraph{The gap this paper addresses.} Opinion pooling has a long statistical
history \citep{genest1986}, and the logarithmic pool is standard. What appears
to be uncommon in football forecasting is its use as a \emph{diagnostic}: fitting
the pooling weight out of sample and reporting it as the measure of whether a
model adds anything to the market. Reporting accuracy alongside a market
accuracy figure, as is customary, does not answer that question, since the two
may be correlated in unknown degree.

\section{An earlier study and its errata}
\label{sec:errata}

The present work began as a correction of \citet{pitcan2018}, an earlier study
by the present author on the same problem, which reported 53.0\% three-way
accuracy over ten Serie A seasons and compared this favourably with betting
markets at 55.3\%. That study contained two framing errors and several
arithmetic ones. We document them here because they are instances of failure
modes the wider literature shares, and because the corrections motivate the
design of what follows.

\paragraph{Framing.} The benchmark was uniform guessing at $33\%$. Over the
present corpus the home side wins $44.1\%$ of matches, draws account for
$26.2\%$ and away wins $29.8\%$, so the majority-class rule alone scores
$44.1\%$; the walk-forward base-rate forecaster used below scores $40.6\%$ on
the test period. Roughly seven percentage points of the reported margin were
therefore not a margin at all. Separately, Bet365 odds were supplied to the
classifiers as features and the resulting accuracy compared against the
bookmaker's, which is the inversion described in Section~\ref{sec:intro}. The
study noted that ``the betting odds did not impact predictions very much'' and
attributed this to the odds being uninformative.

\paragraph{Arithmetic.} The following are recomputed from the confusion
matrices printed in that paper. Both total 218 matches against a stated test
set of 320, a discrepancy the text does not reconcile.

\begin{itemize}
\item Tables 4.1 and 4.2 are both captioned ``Linear SVM''. Table 4.1 has
diagonal $80+29+2=111$ ($50.9\%$), matching the reported Linear SVM figure;
Table 4.2 has diagonal $77+34+3=114$ ($52.3\%$), which is the reported AdaBoost
figure. Table 4.2 is mislabelled.

\item Away wins constitute $60/218 = 27.5\%$ of the test set, so ``never
predict an away win'' scores $72.5\%$. Reported away accuracies are $72.9$
(Gaussian SVM), $69.7$ (Linear SVM), $68.4$ (AdaBoost), $45.9$ (logistic
regression). None materially exceeds the constant rule and three fall below it.

\item Home wins are $92/218$, so ``never predict a home win'' scores
$126/218 = 57.80\%$. The reported Linear SVM home accuracy is $57.8$, which to
three significant figures is that constant rule, implying no home win was ever
predicted.

\item With $n=218$ the standard error of an accuracy near one half is
$\sqrt{0.25/218} \approx 0.034$. The reported multi-class accuracies span
$47.2\%$ to $53.0\%$, i.e.\ under two standard errors end to end, and the
$n=5$ versus $n=7$ comparison moves the three classifiers in inconsistent
directions. No ranking among them is supported.

\item The time-weighted form feature is defined as $\tau'(\{x_1,\dots,x_n\}) =
\frac{1}{n}\sum_i x_i e^{-i}$. A weighted average must be normalised by
$\sum_i w_i$ rather than by $n$; with $n=5$, $\sum_{i=1}^5 e^{-i} = 0.578$, so
the weighted feature carries roughly $1/8.7$ the scale of its unweighted
counterpart, which matters for the kernel and regularised methods used. The
decay is also severe: $63.6\%$ of the weight falls on the most recent match and
the Kish effective sample size \citep{kish1965} is $2.1$ matches, so a nominal
five-match window carries about two matches of information. The decay rate was
asserted rather than fitted.
\end{itemize}

\section{Data}
\label{sec:data}

The corpus comprises nineteen complete Serie A seasons, 2007--08 through
2025--26, totalling $7{,}220$ matches, from \texttt{football-data.co.uk}. Each
record carries the date, the two clubs, the full-time score and outcome, shots,
shots on target and corners for both sides, and bookmaker prices. Bet365
opening prices are available for all seasons; Pinnacle opening and closing
prices from 2012--13 onward. Where available, the Pinnacle closing line is
preferred, as the closing price of a low-margin, high-limit book is the sharpest
widely published forecast.

\begin{figure}[t]
\centering
\includegraphics[width=0.72\textwidth]{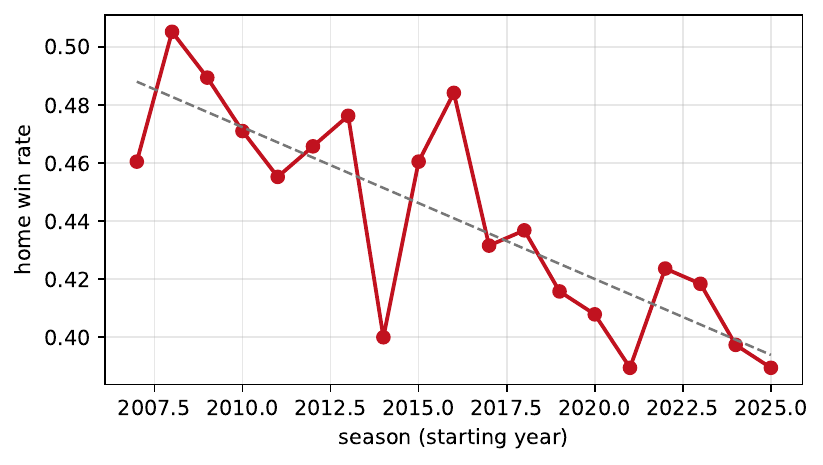}
\caption{Home win rate by season, with a linear trend. The erosion of home
advantage over the sample motivates fitting the decay rate rather than
assuming it.}
\label{fig:home}
\end{figure}

Two features of the corpus bear on modelling choices. First, promotion and
relegation mean that only eight clubs appear in all nineteen seasons while many
appear once or twice, so any rating scheme must handle clubs with little or no
top-flight history. Second, home advantage has eroded substantially over the
period (Figure~\ref{fig:home}): the home win rate falls from $47.5\%$ across
2007--13 to $40.2\%$ across 2023--26, and the ratio of home to away goals from
$1.38$ to $1.14$. A model fitted without time weighting would misstate the
current magnitude of home advantage.

\section{Methods}
\label{sec:methods}

\subsection{Market probabilities}

Let a $k$-outcome book quote decimal odds $o_1,\dots,o_k$ and write
$\pi_i = 1/o_i$ with book sum $\Pi = \sum_j \pi_j > 1$. The excess $\Pi - 1$ is
the bookmaker's margin. Proportional normalisation, $p_i = \pi_i/\Pi$, is known
to leave a favourite--longshot bias. We instead use the model of
\citet{shin1993}, under which quoted prices arise from a bookmaker facing a
proportion $z$ of insider trade; the implied probabilities satisfy
\begin{equation}
p_i(z) \;=\; \frac{\sqrt{z^2 + 4(1-z)\,\pi_i^2/\Pi} \;-\; z}{2(1-z)},
\qquad \sum_{i=1}^{k} p_i(z) = 1 .
\label{eq:shin}
\end{equation}
Differentiating shows $\partial p_i/\partial z < 0$ for all $p_i \in (0,1)$, so
the row sum is strictly decreasing in $z$ and the root is located by bisection.
Over the $17{,}479$ complete books in this corpus the fitted $z$ has median
$0.017$ and maximum $0.061$.

\subsection{Structural model}

Following \citet{dixoncoles1997}, club $j$ carries an attack rating $\alpha_j$
and a defence rating $\delta_j$. For a fixture between home club $h$ and away
club $a$, goals $(X, Y)$ have rates
\begin{equation}
\log \lambda = \gamma + \alpha_h - \delta_a,
\qquad
\log \mu = \alpha_a - \delta_h ,
\end{equation}
where $\gamma$ is the home advantage. The joint mass is
\begin{equation}
\Prob(X = x, Y = y) \;=\; \tau_{\rho}(x, y; \lambda, \mu)\,
\frac{\lambda^{x} e^{-\lambda}}{x!}\,\frac{\mu^{y} e^{-\mu}}{y!} ,
\end{equation}
with the low-score correction
\begin{equation}
\tau_{\rho}(x,y;\lambda,\mu) =
\begin{cases}
1 - \lambda\mu\rho & (x,y) = (0,0),\\
1 + \lambda\rho    & (x,y) = (0,1),\\
1 + \mu\rho        & (x,y) = (1,0),\\
1 - \rho           & (x,y) = (1,1),\\
1                  & \text{otherwise.}
\end{cases}
\end{equation}
This assignment is the unique one conserving total probability: summing
$(\tau_\rho - 1)$ times the independent mass over the four cells gives exactly
zero, whereas transposing the $(0,1)$ and $(1,0)$ terms leaves a residual of
$\rho e^{-\lambda-\mu}(\lambda-\mu)^2$.

Matches are weighted exponentially by age, $w(\Delta t) = \exp(-\xi \Delta t)$
with $\Delta t$ in days before the forecast origin, and parameters maximise the
weighted log-likelihood
\begin{equation}
\ell(\alpha, \delta, \gamma, \rho) = \sum_{m} w(\Delta t_m)
\Big[ \log \tau_\rho(x_m, y_m; \lambda_m, \mu_m)
      + \log \mathrm{Pois}(x_m; \lambda_m)
      + \log \mathrm{Pois}(y_m; \mu_m) \Big].
\end{equation}

\paragraph{Identifiability.} The map $(\alpha, \delta, \gamma) \mapsto
(\alpha + c\mathbf{1},\, \delta + c\mathbf{1},\, \gamma)$ leaves both rates
unchanged, so the parameters are identified only up to this one-dimensional
translation. We impose $\sum_j \alpha_j = 0$, which removes it exactly: the
design matrix over observed ordered pairs has rank one less than its column
count, and the single null direction has non-zero mean in the attack block.
Clubs absent from the training window take the exposure-weighted league mean
rather than an arbitrary constant; because the sum-to-zero constraint applies
to attack but not to defence, defaulting both to zero would assign promoted
clubs a better-than-average defence.

\begin{figure}[t]
\centering
\includegraphics[width=0.62\textwidth]{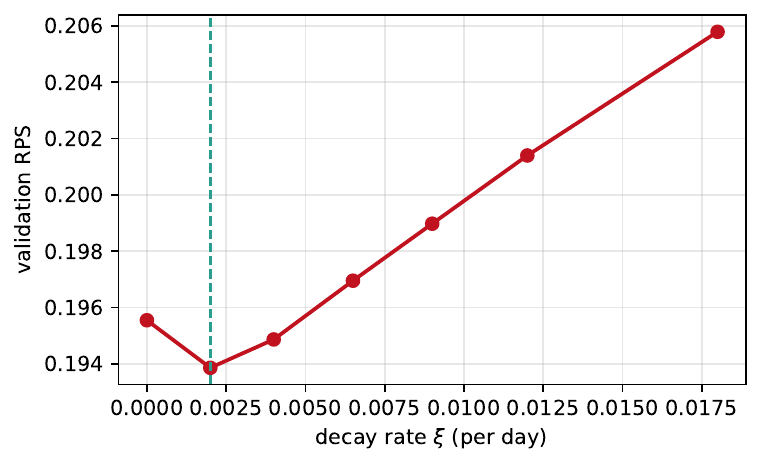}
\caption{Validation RPS across the decay grid. The dashed line marks the
selected rate, $\xi = 0.002$ per day.}
\label{fig:decay}
\end{figure}

\paragraph{Decay rate.} $\xi$ is selected on the validation period by grid
search (Figure~\ref{fig:decay}), giving a clean interior optimum at
$\xi = 0.002$ per day, a half-life of $347$ days---appreciably slower than the
$0.0065$ reported by \citet{dixoncoles1997} for late-1990s English data.

\subsection{A process-based variant}
\label{sec:shots-method}

Goals are a sparse realisation of chance creation, so a rating fitted to them
inherits finishing noise. Expected goals is the standard remedy but is
proprietary, and no freely licensed source covers Serie A back to 2007. We
therefore fit the identical machinery to \emph{shots on target}, the pre-xG
proxy present in the corpus, obtaining shot rates $\lambda^{S}, \mu^{S}$, and
convert to goal rates by a single league-wide finishing factor $\kappa$
estimated on the same window with the same decay weights:
\begin{equation}
\hat\lambda = \kappa_{H}\,\lambda^{S}, \qquad
\hat\mu = \kappa_{A}\,\mu^{S},
\qquad
\kappa_{H} = \frac{\sum_m w_m x_m}{\sum_m w_m s^{H}_m},
\end{equation}
and analogously for $\kappa_A$, where $s^H_m$ denotes home shots on target. The
factor is deliberately league-wide rather than club-specific, since club-level
finishing variation is precisely the noise the substitution is meant to remove.
Fitted values are $\kappa_H = 0.309$, $\kappa_A = 0.317$. The low-score
correction is not carried across, as its parameter would have been estimated
against shot counts; the variant is therefore independent-Poisson on converted
rates, a genuine difference we report rather than conceal.

\subsection{Evaluation protocol}

Because the outcome is ordered, the primary metric is the Ranked Probability
Score \citep{epstein1969,constantinou2012}. For forecast $p$ and realised
indicator $a$ over $r$ ordered categories,
\begin{equation}
\RPS = \frac{1}{r-1}\sum_{i=1}^{r-1}\Bigg(\sum_{j=1}^{i}(p_j - a_j)\Bigg)^{\!2},
\qquad r = 3 ,
\end{equation}
with categories ordered home win, draw, away win. RPS penalises a confident
home forecast more heavily when the away side wins than when the match is
drawn; neither accuracy nor the Brier score makes that distinction. Log loss
and Brier are reported alongside, and accuracy solely for comparability with
prior work.

All forecasts are produced walk-forward: parameters are fitted on matches
strictly preceding an origin, used to forecast the following fortnight, and
refitted at the next origin. Hyperparameters, pooling weights and the
calibration temperature are fixed on 2013--14 to 2018--19 ($n = 2{,}280$); the
test period 2019--20 to 2025--26 ($n = 2{,}660$) is scored once. Every mean
carries a percentile bootstrap interval, and model comparisons use a paired
bootstrap over matches, which exploits the strong correlation induced by
scoring identical fixtures.

\subsection{Incremental information via logarithmic pooling}
\label{sec:pooling}

Let $p^{M}$ denote the de-vigged market forecast and $p^{S}$ the structural
model's. The logarithmic opinion pool \citep{genest1986} with weight
$w \in [0,1]$ is
\begin{equation}
p^{(w)}_i \;\propto\; \big(p^{M}_i\big)^{1-w}\big(p^{S}_i\big)^{w},
\qquad i \in \{H, D, A\},
\label{eq:pool}
\end{equation}
renormalised to sum to one. We fit $\hat w$ by minimising out-of-sample log
loss on the validation period. The logarithmic form is preferred to the linear
one as it is externally Bayesian and operates naturally on log-odds.

The interpretation is the point. If $\hat w$ is indistinguishable from zero,
the market forecast is not improved by any admixture of the model, and the
model carries no information the market has not priced. If $\hat w$ is
materially positive, it does. We propose that this quantity, rather than
accuracy alongside a market accuracy figure, should be the headline diagnostic
in this literature. Equation~\eqref{eq:pool} extends to several models by
raising each to its own weight on the simplex, which we use in
Section~\ref{sec:shots}.

\section{Results}
\label{sec:results}

\subsection{Main comparison}

\begin{table}[t]
\centering
\caption{Test period, 2019--20 to 2025--26 ($n = 2{,}660$). Lower RPS is
better; the interval is a percentile bootstrap on the mean.}
\label{tab:main}
\small
\begin{tabular}{lccccc}
\toprule
Model & RPS & 95\% CI & Log loss & Brier & Accuracy \\
\midrule
Market (Shin de-vigged)     & \textbf{0.1905} & [0.1856, 0.1957] & 0.9620 & 0.5717 & 54.8\% \\
Market $+$ model pool       & 0.1905 & [0.1856, 0.1957] & 0.9620 & 0.5717 & 54.8\% \\
Dixon--Coles (calibrated)   & 0.1972 & [0.1921, 0.2024] & 0.9858 & 0.5862 & 53.4\% \\
Shots on target             & 0.2001 & [0.1959, 0.2042] & 0.9959 & ---    & 52.9\% \\
Base rate (walk-forward)    & 0.2318 & [0.2292, 0.2344] & 1.0846 & 0.6574 & 40.6\% \\
Uniform                     & 0.2340 & [0.2312, 0.2367] & 1.0986 & 0.6667 & 40.6\% \\
\bottomrule
\end{tabular}
\end{table}

Table~\ref{tab:main} reports the central comparison. The market attains RPS
$0.1905$ against the structural model's $0.1972$; the paired bootstrap places
the difference at $+0.0067$ with $95\%$ interval $[0.0046, 0.0088]$. The model
is worse than the market, and decisively so relative to its own sampling error.

Two secondary observations follow. First, the rebuilt model attains $53.4\%$
accuracy against the $53.0\%$ of \citet{pitcan2018}. Nine additional seasons, a
principled likelihood, a fitted decay rate and leak-free validation bought
$0.4$ accuracy points---well inside noise. We read this as evidence that
three-way match accuracy is near a ceiling, and that effort directed at it is
directed at a quantity which has stopped responding. Second, the appropriate
floor is the base-rate forecaster at $40.6\%$, not uniform guessing at $33\%$.

\subsection{Stability across seasons}

\begin{figure}[t]
\centering
\includegraphics[width=0.82\textwidth]{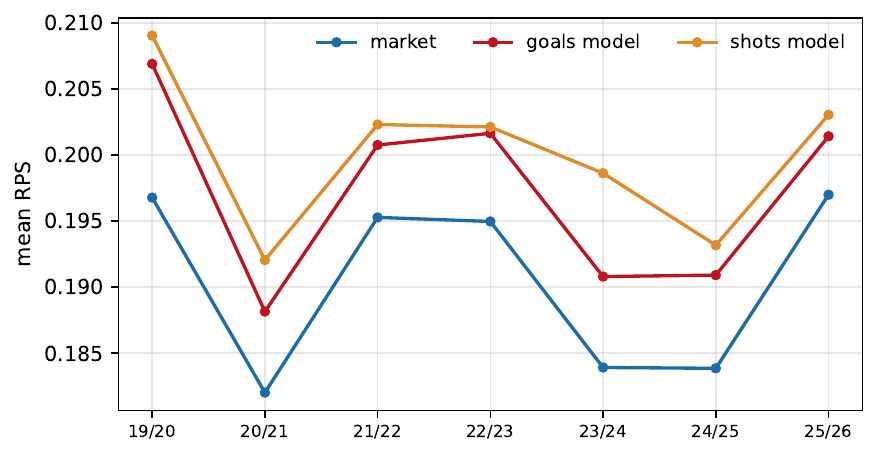}
\caption{Mean RPS by test season. The market is lower in every season; the
two structural models track each other closely.}
\label{fig:seasons}
\end{figure}

A single aggregate gap could be produced by one anomalous season.
Table~\ref{tab:seasons} and Figure~\ref{fig:seasons} show it is not: the market
attains lower RPS in all seven test seasons, with a gap ranging from $0.0044$
to $0.0101$ and no season in which the ordering reverses.

\begin{table}[t]
\centering
\caption{Season-by-season mean RPS on the test period ($n = 380$ per season).
``Gap'' is the goals model minus the market.}
\label{tab:seasons}
\small
\begin{tabular}{lccccc}
\toprule
Season & $n$ & Market & Goals model & Shots model & Gap \\
\midrule
2019--20 & 380 & 0.1968 & 0.2069 & 0.2090 & $+0.0101$ \\
2020--21 & 380 & 0.1820 & 0.1881 & 0.1920 & $+0.0061$ \\
2021--22 & 380 & 0.1953 & 0.2008 & 0.2023 & $+0.0055$ \\
2022--23 & 380 & 0.1950 & 0.2016 & 0.2021 & $+0.0067$ \\
2023--24 & 380 & 0.1839 & 0.1908 & 0.1986 & $+0.0069$ \\
2024--25 & 380 & 0.1838 & 0.1909 & 0.1932 & $+0.0071$ \\
2025--26 & 380 & 0.1970 & 0.2014 & 0.2030 & $+0.0044$ \\
\bottomrule
\end{tabular}
\end{table}

\subsection{The pooling weight}
\label{sec:pool-result}

\begin{figure}[t]
\centering
\includegraphics[width=0.66\textwidth]{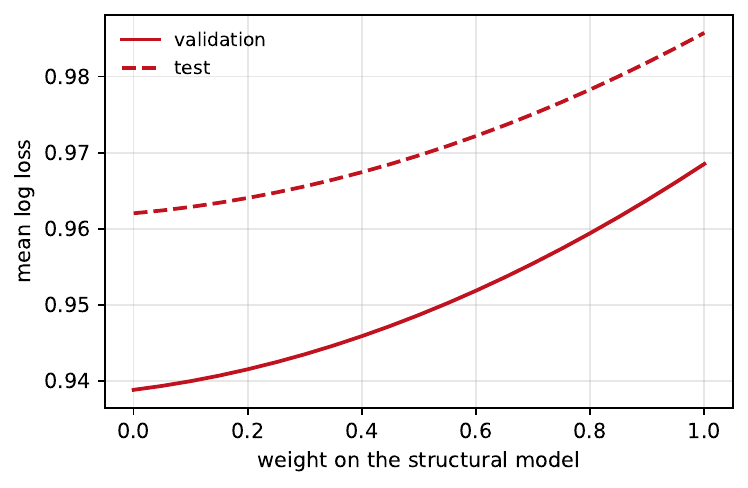}
\caption{Mean log loss of the pooled forecast as a function of the weight
placed on the structural model. Monotone increasing on both periods, with the
minimum at $w = 0$.}
\label{fig:profile}
\end{figure}

Fitted on the validation period, the weight $\hat w$ on the structural model in
the pool of Equation~\eqref{eq:pool} is $0.000$. Because this lies on the
boundary of the admissible interval, a fitted value alone cannot distinguish a
genuine null contribution from an optimiser terminating at a bound. We
therefore trace the entire profile (Figure~\ref{fig:profile}): log loss
increases monotonically in $w$ across $[0,1]$, on the validation period and on
the test period alike. The argument minimum is $w = 0$ in both cases. The
result is not an artefact of the fitting procedure, and it does not depend on
the validation/test split---even an analyst who improperly fitted the weight on
the test data would obtain zero.

The interval $[0,1]$ is the admissible range for a pool: outside it the
combination is no longer a mixture of two opinions. A sceptical reader may
nonetheless ask whether the unconstrained minimum lies below zero, which would
say something stronger than that the model is uninformative---that it is
anti-informative given the price. Extending the profile to $w \in [-1,1]$, the
validation minimum is interior, at $\hat w = -0.225$, with the loss rising
again for weights more negative than that. Freezing that weight and carrying it
to the test period improves mean log loss by $0.00062$ ($95\%$ CI
$[-0.00105, 0.00231]$) and mean RPS by $0.00021$ ($95\%$ CI
$[-0.00029, 0.00071]$). Both intervals contain zero, so the tilt is not a
reliable out-of-sample improvement.

We read the sign rather than the magnitude. A negative weight divides the
market forecast by the structural one, which sharpens it, and
Section~\ref{sec:calibration} shows the structural model to be the flatter of
the two. The unconstrained optimum is therefore acting as a temperature
correction on the price, not as a channel through which the model contributes
information.

\subsection{Is the target too noisy?}
\label{sec:shots}

The natural objection is that the model failed because goals are noisy rather
than because the method is inadequate. Table~\ref{tab:pool} addresses this
using the multi-model extension of Equation~\eqref{eq:pool}.

\begin{table}[t]
\centering
\caption{Pool weights fitted on the validation period.}
\label{tab:pool}
\small
\begin{tabular}{lccc}
\toprule
Pool & Market & Goals model & Shots model \\
\midrule
Market $+$ shots           & 1.00 & ---  & 0.00 \\
Goals $+$ shots            & ---  & 0.65 & \textbf{0.35} \\
Market $+$ goals $+$ shots & 1.00 & 0.00 & 0.00 \\
\bottomrule
\end{tabular}
\end{table}

Against the goals model alone, the shots variant earns weight $0.35$: it
demonstrably carries information the goals model does not, so the negative
result of Section~\ref{sec:pool-result} is not an artefact of target choice.
Against the market, the same signal earns $0.00$; and in the three-way pool
both structural models collapse to zero simultaneously. Two model families
built on different targets, each holding information the other lacks, and the
closing price has already absorbed both. We regard this as the strongest form
of the negative result available without licensed event data, and note that
expected goals---a refinement of the same process-based idea---would most
plausibly sharpen the shots variant rather than reverse its relationship to the
market.

\subsection{Calibration and discrimination}
\label{sec:calibration}

\begin{figure}[t]
\centering
\includegraphics[width=0.52\textwidth]{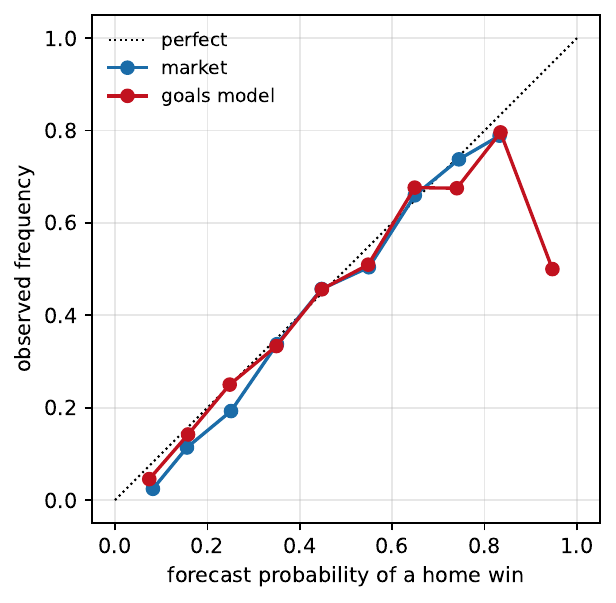}
\caption{Home-win reliability curves on the test period. Both forecasters track
the diagonal closely; the market's advantage lies in sharpness rather than
calibration.}
\label{fig:calib}
\end{figure}

Table~\ref{tab:calib} and Figure~\ref{fig:calib} separate two properties that
an accuracy figure conflates. Regressing the outcome indicator on the forecast
log-odds gives a calibration slope of $0.995$ for the structural model against
$1.103$ for the market, with essentially identical expected calibration error.
A slope near one indicates the stated probabilities can be taken at face value;
the market's slope above one indicates mild under-confidence on this margin.
Yet the market is decisively sharper, which is what drives its RPS advantage.
The fitted recalibration temperature for the structural model is $0.99$, i.e.\
essentially no correction is warranted.

\begin{table}[t]
\centering
\caption{Calibration on the test period. ECE is the count-weighted mean
absolute gap between stated and observed frequency over ten bins; slope and
intercept are from a logistic regression of the home-win indicator on the
forecast log-odds.}
\label{tab:calib}
\small
\begin{tabular}{lccccc}
\toprule
Model & ECE (H) & ECE (D) & ECE (A) & Slope (H) & Intercept (H) \\
\midrule
Dixon--Coles            & 0.0208 & 0.0151 & 0.0270 & \textbf{0.995} & $-0.054$ \\
Market (Shin de-vigged) & 0.0285 & 0.0095 & 0.0253 & 1.103 & $-0.090$ \\
Base rate               & 0.0164 & 0.0102 & 0.0062 & 0.191 & $-0.321$ \\
\bottomrule
\end{tabular}
\end{table}

The practical implication is that ``the market beats the model'' and ``the
model's probabilities cannot be trusted'' are different statements, and only
the first holds here. A well-calibrated but less sharp forecast remains usable
for any decision that consumes the probability itself, which is the basis for
Section~\ref{sec:leverage}. The base-rate row illustrates the converse
pathology: a forecast can be near-perfectly calibrated while carrying almost no
discriminatory content, its slope of $0.191$ reflecting the absence of any
signal to be calibrated.

\section{Match leverage}
\label{sec:leverage}

\subsection{Definition}

If a structural model cannot improve on a closing price for match outcomes, the
product worth building is not a match-outcome forecast, which a club can obtain
for free. What the market does not supply is the quantity that governs a club's
decisions: how much a particular fixture matters to its season.

Let $\Omega$ be a season objective expressed as a set of admissible final league
positions---survival, European qualification, the championship. For a remaining
fixture $m$, define
\begin{equation}
L_m \;=\; \Prob\big(\Omega \,\big|\, \text{club wins } m\big)
       \;-\; \Prob\big(\Omega \,\big|\, \text{club loses } m\big).
\label{eq:leverage}
\end{equation}

\subsection{Estimation}

Let $S$ denote the scoreline of fixture $m$ and $R$ the scorelines of all other
remaining fixtures. The objective indicator is a deterministic function
$f(S, R)$: it depends on $m$ through the points awarded to both clubs, and on
every other result through the final ordering. If the simulator draws
$S \perp R$, then for any result set $W$,
\begin{equation}
\Prob\big(f(S,R) = 1 \,\big|\, S \in W\big)
= \sum_{s \in W} \Prob(S = s \mid S \in W)\; \Prob\big(f(s, R) = 1\big),
\end{equation}
which is exactly the interventional quantity obtained by forcing $m$ to a result
drawn from the model's own conditional distribution given $W$ and resampling the
remainder. Conditioning on $S$ cannot induce selection on $R$ precisely because
the two are independent; the opponent's points \emph{from} $m$ are part of $S$
and are conditioned on, not marginalised. Hence a single simulation run supports
the estimation of $L_m$ for every fixture simultaneously, by post-hoc
conditioning.

We verified this equivalence in two ways: by exhaustive enumeration of all
$59{,}049$ joint outcomes of a reduced four-club league, where post-hoc and
forced-result computations agree to machine precision; and by direct comparison
against forced-result simulation at $10^{6}$ replications on the applied case
below, where no fixture differs beyond Monte Carlo error (maximum $|z| = 1.17$
across twelve fixtures).

In practice we simulate the remainder of the season $20{,}000$ times, drawing
complete scorelines from each fixture's joint distribution so that goal
difference resolves naturally, and rank the final table by points, then goal
difference, then goals scored. (Serie A resolves equal points on head-to-head
record first; exact ties affecting the reported band arise in under $0.01\%$ of
simulations.)

\subsection{Application}

\begin{table}[t]
\centering
\caption{ACF Fiorentina, simulated from 1 March 2026 with the club 16th on 24
points and twelve fixtures remaining. Objective: survival, i.e.\ a final
position of 17th or better. Monte Carlo standard error on each leverage
estimate is below $0.005$.}
\label{tab:leverage}
\small
\begin{tabular}{llcccc}
\toprule
Fixture & Venue & $\Prob(\Omega \mid W)$ & $\Prob(\Omega \mid D)$ & $\Prob(\Omega \mid L)$ & $L_m$ \\
\midrule
Lecce      & A & 0.987 & 0.959 & 0.914 & \textbf{0.073} \\
Cremonese  & A & 0.986 & 0.957 & 0.922 & 0.063 \\
Verona     & A & 0.982 & 0.952 & 0.927 & 0.055 \\
Genoa      & H & 0.984 & 0.962 & 0.932 & 0.053 \\
Udinese    & A & 0.985 & 0.961 & 0.933 & 0.052 \\
Sassuolo   & H & 0.982 & 0.956 & 0.932 & 0.050 \\
Lazio      & H & 0.988 & 0.960 & 0.942 & 0.046 \\
Parma      & H & 0.981 & 0.955 & 0.936 & 0.045 \\
Atalanta   & H & 0.990 & 0.966 & 0.949 & 0.041 \\
Juventus   & A & 0.988 & 0.971 & 0.951 & 0.037 \\
Roma       & A & 0.988 & 0.968 & 0.952 & 0.036 \\
Inter      & H & 0.988 & 0.971 & 0.955 & \textbf{0.033} \\
\bottomrule
\end{tabular}
\end{table}

Table~\ref{tab:leverage} reports the result. The ordering inverts intuition.
The fixtures moving survival most are those against direct relegation rivals;
those against the leading clubs move it least, because defeat is already
expected in both branches and therefore already priced into the season
distribution. Losing at home to Inter---the eventual champions, and the
season's marquee fixture---costs $3.3$ percentage points of survival
probability; losing away at Lecce costs $7.3$. The away fixture carries
$2.25\times$ the leverage of the home one, a gap of approximately eight Monte
Carlo standard errors, so the ordering is not simulation noise.

This is a rotation and load-management statement expressed in probability: the
Inter fixture is where a fatigued first-choice player is cheapest to rest, and
Lecce away is where he is most expensive. No match-outcome forecast conveys it.

As a coherence check rather than a claim of predictive success, the simulation
projected $41.4$ expected final points against an actual $42$ and assigned
survival probability $96.4\%$; Fiorentina finished 15th. A single realisation
is weak evidence, but it establishes that the season aggregation is not grossly
miscalibrated, which is the minimum precondition for acting on a leverage
ranking.

\section{Discussion}
\label{sec:discussion}

\paragraph{What the negative result does and does not say.} It does not say that
structural models are useless, nor that the market is efficient in a strong
sense. It says that over this league and period, for two structurally different
signals, the marginal information relative to a de-vigged closing price is
indistinguishable from zero at the resolution afforded by $2{,}660$ matches. A
model with genuinely private inputs---player availability, training-load data,
tactical intelligence---is not tested here and is exactly where we would expect
remaining edge to lie.

\paragraph{Implications for reporting practice.} We would encourage two changes.
First, report the base rate as the floor; uniform guessing is not a meaningful
benchmark for an outcome with a $44\%$ majority class. Second, where market
prices are available, report the fitted pooling weight against them. It is a
single number, it is cheap to compute, and it answers the question a reader
actually has. An accuracy figure quoted beside a market accuracy figure does
not, because the two forecasts may be arbitrarily correlated.

\paragraph{Calibration as the usable residual.} The finding that the model is
better calibrated but less sharp than the market has a constructive reading. A
club cannot profitably trade on this model, but it can aggregate it: the season
simulation of Section~\ref{sec:leverage} consumes probabilities directly, and
what such aggregation requires is that stated probabilities mean what they say,
not that they be maximally sharp. This is the sense in which the useful product
sits on top of a calibrated forecast rather than in the forecast itself.

\paragraph{On the ceiling.} That a decade of methodological improvement and
nine additional seasons moved accuracy by $0.4$ points is worth stating
plainly. Under a scoring rule sensitive to the full distribution the
improvement is likewise small. If the goal is a better outcome forecast, the
binding constraint is information, not method.

\section{Limitations}
\label{sec:limits}

\textbf{No player-level information.} No injuries, suspensions, rest days or
transfers enter the model. This is the largest gap relative to a club's internal
information set and the most plausible location of remaining edge.

\textbf{No expected goals.} Shots on target weight a tap-in and a speculative
long-range effort equally. Licensed event data would sharpen the process-based
variant; the results of Section~\ref{sec:shots} suggest refinement rather than
reversal, but this is not established.

\textbf{The benchmark embeds late information.} The closing price incorporates
team news the model never sees, so some portion of the $0.0067$ RPS gap is
information asymmetry rather than modelling deficiency. This works against the
model and in favour of the reported conclusion.

\textbf{One league, one period.} Serie A over nineteen seasons. Market
efficiency need not transfer to leagues with thinner betting markets, and the
erosion of home advantage documented in Section~\ref{sec:data} shows the
data-generating process is not stationary.

\textbf{Leverage is single-objective.} Equation~\eqref{eq:leverage} conditions
on one position band. A club trading survival against a cup run requires a
utility function over outcomes, not a probability difference.

\section{Reproducibility and verification}
\label{sec:repro}

All results derive from public data and a versioned Python package with 126
tests. Beyond unit testing, four components---the margin-removal solver, the
likelihood and its identifiability, the walk-forward harness, and the leverage
estimator---were subjected to independent adversarial audit by parties who did
not write them. That audit confirmed the absence of look-ahead bias by positive
control (an injected one-line boundary error improves test RPS by $0.0123$ and
reverses the sign of the market comparison, a signature the reported pipeline
does not exhibit) and identified four defects, all corrected before the results
reported here were generated. Two of the corrected defects were tests that
asserted nothing: one passed against a deliberately leaky harness, the other
verified only that probabilities summed to one. We mention this because the
negative result reported here is of a kind that a coding error can easily
manufacture, and readers are entitled to know what was done to exclude that.

\textbf{Platform sensitivity.} The figures reported here reproduce exactly on
the machine that generated them: two independent runs of the full pipeline
agree bit for bit. They do not reproduce exactly across machines. No random
seed enters the fit---the likelihood is optimised by L-BFGS-B from a
deterministic starting point with finite-difference gradients---so the only
source of variation is floating-point reduction order, which differs between
linear-algebra builds. Rerunning on different hardware moved the converged
parameters in the fifth decimal, which propagated to the calibration
temperature ($0.99011$ against $0.99009$) and, on the test period, flipped one
match of $2{,}660$ from incorrect to correct. Every figure in
Tables~\ref{tab:main}, \ref{tab:seasons}, \ref{tab:pool} and
\ref{tab:leverage} survives at the precision printed; two expected-calibration
entries in Table~\ref{tab:calib} move by $0.0003$. A reader reproducing the
pipeline should expect agreement to three decimal places rather than exactly.

Code, data pipeline and figures: \url{https://github.com/pitcany/seriea-leverage}.

\section{Conclusion}

Over nineteen Serie A seasons, a well-specified Dixon--Coles model earns a
logarithmic pooling weight of zero against a Shin de-vigged closing price, and
so does a process-based variant fitted to shots on target---though the latter
earns weight $0.35$ against the former, establishing that both carry real and
partially distinct information. The market has already absorbed both. The model
is nonetheless better calibrated than the market while being less sharp, which
makes it usable for aggregation even where it is unusable for prediction. We
define match leverage as one such aggregate and show that it inverts intuitive
fixture importance: for a club fighting relegation, an away match against a
fellow struggler was worth more than twice as much as hosting the eventual
champions.

\subsection*{Data and code availability}

Match data are publicly available from \texttt{football-data.co.uk}. All code,
including the exact scripts producing every table and figure here, is at
\url{https://github.com/pitcany/seriea-leverage}.

\end{document}